\documentclass[pdflatex,prl,twocolumn,showpacs,superscriptaddress]{revtex4-1}
\usepackage{bm,amsmath,amssymb}
\usepackage{graphicx}
\usepackage{subfigure}
\usepackage{color}
\usepackage[normalem]{ulem}  % \sout{old text} for strikeout
\usepackage{hyperref}
\usepackage{multirow}
\newcommand \beq {\begin{equation}}
\newcommand \eeq {\end{equation}}
\newcommand \beqa {\begin{eqnarray}}
\newcommand \eeqa {\end{eqnarray}}
\newcommand \hmu {\hat{\mu}}
\newcommand \bmu {\bar{\mu}}
\def\lsim{\raise0.3ex\hbox{$<$\kern-0.75em\raise-1.1ex\hbox{$\sim$}}}
\def\gsim{\raise0.3ex\hbox{$>$\kern-0.75em\raise-1.1ex\hbox{$\sim$}}}
\usepackage[normalem]{ulem}  % \sout{old text} for strikeout

\begin{document}
%\linenumbers
%\begin{frontmatter}

%------------------------------------------------------------------------------------
%       title
%------------------------------------------------------------------------------------

\title{The curvature of the freeze-out line in heavy ion collisions
}
\author{
A. Bazavov}
\affiliation{Department of Physics and Astronomy, University of Iowa, 
Iowa City, Iowa 52240, USA}
\author{
H.-T. Ding}
\affiliation{Key Laboratory of Quark \& Lepton Physics (MOE) and Institute
of Particle Physics, Central China Normal University, Wuhan 430079, China}
\author{
P. Hegde} 
\affiliation{Key Laboratory of Quark \& Lepton Physics (MOE) and Institute
of Particle Physics, Central China Normal University, Wuhan 430079, China}
\author{
O. Kaczmarek}
\affiliation{
Fakult\"at f\"ur Physik, Universit\"at Bielefeld, D-33615 Bielefeld,
Germany}
\author{
F. Karsch}
\affiliation{
Fakult\"at f\"ur Physik, Universit\"at Bielefeld, D-33615 Bielefeld,
Germany}
\affiliation{
Physics Department, Brookhaven National Laboratory, Upton, NY 11973, USA}
\author{
E. Laermann} 
\affiliation{
Fakult\"at f\"ur Physik, Universit\"at Bielefeld, D-33615 Bielefeld,
Germany} 
\author{
Swagato Mukherjee} 
\affiliation{
Physics Department, Brookhaven National Laboratory, Upton, NY 11973, USA}
\author{H. Ohno}
\affiliation{
Physics Department, Brookhaven National Laboratory, Upton, NY 11973, USA}
\affiliation{Center for Computational Sciences, University of Tsukuba,
Tsukuba, Ibaraki 305-8577, Japan}
\author{
P. Petreczky}
\affiliation{
Physics Department, Brookhaven National Laboratory, Upton, NY 11973, USA}
\author{
C. Schmidt} 
\affiliation{
Fakult\"at f\"ur Physik, Universit\"at Bielefeld, D-33615 Bielefeld,
Germany}
\author{S. Sharma}
\affiliation{
Physics Department, Brookhaven National Laboratory, Upton, NY 11973, USA}
\author{
W. Soeldner}
\affiliation{
Institut f\"ur Theoretische Physik, Universit\"at Regensburg,
D-93040 Regensburg, Germany}
\author{
M. Wagner}
\affiliation{Physics Department, Indiana University, Bloomington, IN 47405,
USA}

\date{\today}

\begin{abstract}
We calculate the mean and variance of net-baryon number and net-electric charge distributions
from Quantum Chromodynamics (QCD) using a next-to-leading order Taylor expansion in terms of temperature and chemical potentials. 
We compare these
expansions with experimental data from STAR and PHENIX, determine the freeze-out temperature
in the limit of vanishing baryon chemical potential, 
and, for the first time, constrain the curvature
of the freeze-out line through a direct comparison between experimental data on net-charge
fluctuations and a QCD calculation. We obtain a bound on the curvature coefficient, 
$\kappa_2^f < 0.011$, that is compatible with lattice QCD results on the curvature of the QCD transition line. 
\end{abstract}

\maketitle

\section{Introduction}
Heavy ion collisions at varying beam energies are performed at the Relativistic
Heavy Ion Collider (RHIC) with the goal to probe properties of strong-interaction 
matter at different temperatures and baryon chemical potentials. This Beam Energy
Scan program aims for exploring physics in the vicinity of the
pseudo-critical line of the transition from hadronic matter to the quark-gluon
plasma. The hope is to find evidence for the existence of a critical point 
\cite{critical} in
the phase diagram of strong-interaction matter that marks the end of a line
of first order phase transitions. 

With decreasing beam energy, $\sqrt{s_{_{NN}}}$, more baryons from the 
incident nuclei are stopped leading to an increase in the baryon number density and thus 
also to an increase of the baryon chemical potential in a central rapidity window.
The dense matter created in these collisions is expected to reach local 
thermal equilibrium quickly. It subsequently expands and cools down.
Eventually hadrons start to form again. This hadronization is characterized 
by a temperature ($T$) and baryon chemical potential ($\mu_B$) that varies with
$\sqrt{s_{_{NN}}}$.
Shortly thereafter inelastic interactions among the hadrons cease to occur and
different particle species ''freeze-out''. This so-called chemical
freeze-out is characterized by a set of freeze-out
parameters, $(T_f,\mu_B^f)$. With varying $\sqrt{s_{_{NN}}}$ they  
map out a line in the $T$-$\mu_B$ plane, called the freeze-out line, 
$T_f(\mu_B)$. An 
important question is how close the freeze-out line is to 
the pseudo-critical line, $T_c(\mu_B)$ \cite{Kaczmarek:2011zz,Endrodi:2011gv}, 
characterizing the crossover transition of QCD. The proximity
of both lines in the phase diagram is a pre-requisite for being able to
explore critical behavior in the vicinity of a possibly existing critical 
point, by analyzing observables that can probe physics on the freeze-out line. 

Unlike the crossover line, $T_c(\mu_B)$, for which the transition temperature
at $\mu_B=0$ \cite{Aoki:2009sc,Tc} and the leading order correction 
at small $\mu_B^2$ \cite{Kaczmarek:2011zz,Endrodi:2011gv} have been determined 
in lattice QCD calculations, 
the parametrization of the freeze-out line, $T_f(\mu_B)$, is not 
given in terms of fundamental parameters of QCD. The line characterizes
the expanding medium formed in heavy ion collisions. 
Freeze-out parameters have been determined in a wide range of $\sqrt{s_{_{NN}}}$ 
by comparing experimental data on particle yields with statistical hadronization 
models (HRG models). Parametrizations of the freeze-out line $T_f(\mu_B)$
have been extracted from such analyses \cite{Cleymans:2005xv,Andronic:2005yp}.
Physically motivated ans\"atze, such as identifying $T_f(\mu_B)$ with a
line of constant energy per particle \cite{Cleymans:2005xv}, naturally lead to the 
expectation that $T_f(\mu_B)$ is a function of $\mu_B^2$ that decreases with increasing 
$\mu_B$. In fact, a simple ansatz 
\cite{Cleymans:2005xv},
\begin{equation}
T_f(\mu_B) = T_{f,0} \left( 1 - \kappa_2^f \bmu_B^2 -\kappa_4^f \bmu_B^4 \right) \; ,
\label{Tf}
\end{equation}
with $\bmu_B\equiv \mu_B/T_{f,0}$,
provides a good parametrization in the entire energy range probed in heavy ion
collisions. 
Using this ansatz for the comparison of HRG model calculations with experimental 
data obtained for a wide range of energies $\sqrt{s_{_{NN}}}$ gave 
$\kappa_2^f= 0.023(3)$ \cite{Cleymans:2005xv} for the 
curvature coefficient. However, a recently performed refined hadronization 
model analysis suggests a weaker energy dependence of the freeze-out line
\cite{Becattini}. 
Moreover, the attempt to capture more accurately the behavior of freeze-out
temperatures at large $\sqrt{s_{_{NN}}}$, i.e. small $\mu_B$, lead to a phenomenological 
parametrization \cite{Andronic:2005yp} that does not even have a power-like dependence  
on $\mu_B$, 
i.e. $T_f(\mu_B) - T_{f,0} \sim {\rm exp}(-a/\mu_B)$ which 
favors $\kappa_2^f\simeq 0$.

In this paper we improve over the current situation by outlining a 
procedure to determine the ${\cal O}(\mu_B^2)$ coefficient of the 
freeze-out line from measurements of the mean and variance of net-electric 
charge and net-proton number distributions and, for the first time, 
illustrate how this can
be done by comparing experimental data obtained at large $\sqrt{s_{_{NN}}}$ 
directly with a QCD 
calculation. This puts the determination of the curvatures of the freeze-out line on a par with that of the QCD transition line.

\section{Ratio on charge fluctuations on the freeze-out line}
The mean, $M_X\equiv \chi_1^X(T,\mu)$, and variance, $\sigma_X^2\equiv \chi_2^X(T,\mu)$, 
of net-electric charge ($X=Q$) and net-baryon number ($X=B$)  
distributions are obtained 
as functions of $T$ and
$\mu\equiv (\mu_B,\mu_Q,\mu_S)$ by taking 
derivatives of the QCD pressure with respect to $\hmu_X\equiv \mu_X/T$  
\begin{equation}
\chi_n^X (T,\mu) = \frac{\partial^n P/T^4}{\partial \hmu_X^n} \; ,\;
X=B,\ Q,\ S \; .
\label{chiX}
\end{equation}
The ratios of mean and variance, 
\begin{equation}
R_{12}^X(T,\mu)\equiv \frac{M_X}{\sigma_X^2} =
\frac{\chi_1^X(T,\mu)}{\chi_2^X(T,\mu)} 
 \ ,
\label{chiX}
\end{equation}
can be analyzed in heavy ion experiments.
Implementing the constraints $M_S=0$ and $M_Q/M_B=r$, which are
appropriate for the initial conditions met in such collisions,
the ratios $R_{12}^B$ and $R_{12}^Q$
become functions of $T$ and $\mu_B$ only and 
$\Sigma_r^{QB}\equiv R_{12}^Q/R_{12}^B =r \sigma_B^2/\sigma_Q^2$.
In leading order (LO) Taylor 
expansion $\Sigma_r^{QB}$ is independent of $\hmu_B$, while the ratios 
$R_{12}^B$  and $R_{12}^Q$ depend linearly on $\hmu_B$,
$R_{12}^X = R_{12}^{X,1} \hmu_B + {\cal O}(\hmu_B^3)$. They therefore provide 
mutually independent information that can be used to extract $T_f(\mu_B)$ up to
${\cal O}(\mu_B^2)$ \cite{Karsch:2012wm}.
 
In order to determine the freeze-out temperature at $\mu_B=0$ and 
the curvature of the freeze-out line, 
we consider a next-to-leading order Taylor expansion of 
$\Sigma_r^{QB}$ in terms of $T$ and $\hmu_B$ around the 
point ($T_{f,0},\mu_B=0$) with $\hmu_S$ and $\hmu_Q$ being implicit 
functions of $T$ and $\hmu_B$. Using Eq.~\ref{Tf} as a parametrization
of the freeze-out line, we find in next-to-leading order (NLO),
\begin{equation}
\hspace*{-0.1cm}\Sigma_r^{QB} =\Sigma_r^{QB,0} + \left( \hspace*{-0.1cm} \left. 
\Sigma_r^{QB,2} -\kappa_2^f T_{f,0}
\frac{{\rm d} \Sigma_r^{QB,0}}{{\rm d} T} \right|_{T_{f,0}}       
\right) \hmu_B^2 \ .
\label{NLO}
\end{equation}
The LO expansion coefficient is easily related to the quadratic
fluctuations of net-electric charge and net-baryon number,
$\Sigma_r^{QB,0} = r \chi_2^B(T)/\chi_2^Q(T)$
at zero $\hmu_B$. 
The NLO expansion 
coefficient $\Sigma_r^{QB,2}$ depends on fourth order cumulants, 
which also can be calculated in a lattice QCD calculation
at vanishing $\mu_B$. An explicit expression for $\Sigma_r^{QB,2}$ will 
be given elsewhere \cite{next}. 

In order to facilitate a comparison with the experimental data it is of
advantage to eliminate $\hmu_B$ from
Eq.~\ref{NLO} in favor of observables that are accessible to
experiments and QCD calculations. For a consistent treatment of the NLO
result, Eq.~\ref{NLO}, it suffices to use the LO relation between
$\hmu_B$ and the ratio $R_{12}^B$, i.e. $\hmu_B= R_{12}^B/R_{12}^{B,1}$.
The LO expansion coefficient $R_{12}^{B,1}$ has been evaluated before
and continuum extrapolated results for $m_l=m_s/20$ obtained
with the HISQ action have been shown in \cite{BIBNL}.

After replacing $\hmu_B$ in favor of $R_{12}^B$,  
the NLO Taylor expansion of $\Sigma_r^{QB}$, introduced in
Eq.~\ref{NLO},, becomes
\begin{equation}
\Sigma_r^{QB} = a_{12}\left( 1 +c_{12} \left( R_{12}^{B}\right)^2 \right) 
+{\cal O} \left( \left( R_{12}^{B}\right)^4\right)\; ,
\label{relation}
\end{equation}
where $a_{12}(T)\equiv \Sigma_r^{QB,0}$. 
The coefficient of the quadratic correction, $c_{12}$, depends on the
parametrization of the freeze-out line 
and needs to be determined at $T_{f,0}$,
\begin{eqnarray}
\hspace*{-0.1cm}
c_{12}(T_{f,0},\kappa_2^f) &=& c_{12}^0(T_{f,0}) -\kappa_2^f D_{12}(T_{f,0})
\label{c12a}
\end{eqnarray}
with
\begin{eqnarray}
c_{12}^{0}(T) &=& \left( \frac{1}{R_{12}^{B,1}}\right)^2 
\frac{\Sigma_r^{QB,2}}{\Sigma_r^{QB,0}}\; , 
\nonumber \\
D_{12}(T) &=& \left( \frac{1}{R_{12}^{B,1}}\right)^2  
T \frac{{\rm d} \ln \Sigma_r^{QB,0}}{{\rm d} T}  \ .
\label{c12b}
\end{eqnarray}

\begin{figure*}[t]
\begin{center}
\includegraphics[width=174mm]{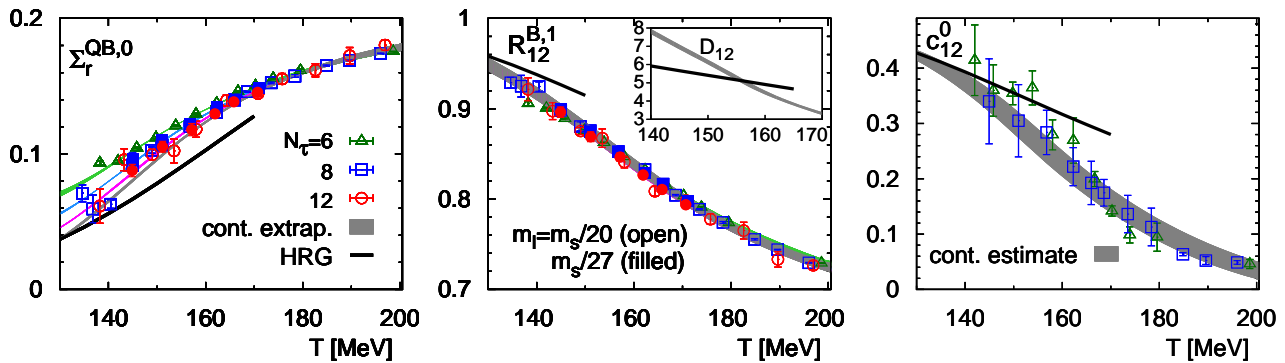}
\caption{Temperature dependence of the LO expansion coefficients 
of the ratio $\Sigma_r^{QB}$ (left) and $R_{12}^B$ (middle) 
and the NLO coefficient $c_{12}^0$ (right) introduced in Eq.~\ref{c12b}.
The NLO coefficient $D_{12}$ (Eq.~\ref{c12b}) is shown as an insertion
in the middle panel. 
Shown are data from calculations with the HISQ action on lattices of
size $N_\sigma^3\times N_\tau$, with $N_\sigma=4 N_\tau$.
Bands show continuum extrapolations for the LO observables and
continuum estimates for the NLO observable $c_{12}^0$ (see text).
}
\label{fig:chi2QB}
\end{center}
\end{figure*}

In 
Fig.~\ref{fig:chi2QB} we show results for 
$\Sigma_r^{QB,0}$, $R_{12}^{B,1}$ and $c_{12}^0$
obtained in lattice QCD calculations 
with the (2+1)-flavor HISQ action \cite{Follana} using a physical value
of the strange quark mass and two sets of light quark
masses, $m_l=m_s/20,\ m_s/27$, which in the continuum limit
correspond to pion mass values, $m_\pi\simeq 160$~MeV and 140~MeV, 
respectively. 
Here we used $r\simeq 0.4$ which is appropriate for describing
colliding gold or lead 
nuclei. Further details on the simulation parameters can be found in
\cite{hotQCD}. We note that quark mass effects are small
for the observables under consideration. 

The continuum extrapolated results for 
$\Sigma_r^{QB,0}$ and
$R_{12}^{B,1}$ 
in the left and middle panels of 
Fig.~\ref{fig:chi2QB}   
were obtained by
performing cubic spline fits to all $m_l=m_s/20$ data, with $1/N_{\tau}^2$ dependence for
the spline coefficients and for the varying locations of three knots. These 
fits were performed over many bootstrap samples drawn from the Gaussian errors
of data points
and have been constrained to agree with HRG at $T=130$~MeV within 10\%.
The final continuum results were obtained from mean values and errors
of these bootstrapped fit results weighted by the quality of the fits given by the
Akaike information criteria. The continuum extrapolations are
consistent with our earlier results \cite{hotQCD}. However, statistical 
errors are reduced considerably.

The parameter
$D_{12}$ is obtained from the continuum extrapolated results for
$R_{12}^{B,1}$ and $\Sigma_r^{QB,0}$.  It is shown as an insertion in the middle panel
of Fig.~\ref{fig:chi2QB}.  The coefficient $c_{12}^0$ receives contributions from
fourth order cumulants and thus is more difficult to extract.  At present we have
calculated it for two lattice spacings and therefore only provide an estimate for its
continuum limit.  This is shown in the right panel of Fig.~\ref{fig:chi2QB}. 
For the continuum estimate of $c_{12}^0$ we followed an
identical procedure of the cubic spline fits outlined above 
but with only two knots.

\begin{figure}[t]
\begin{center}
\includegraphics[width=68mm]{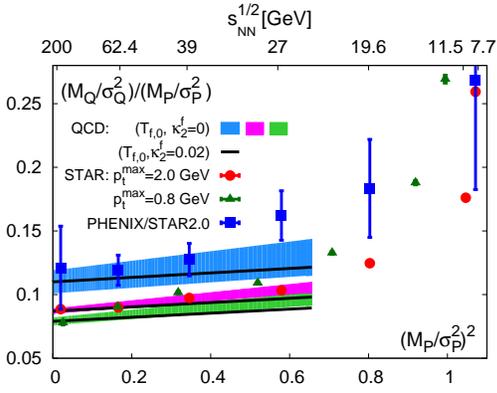}
\caption{The ratio of ratios of mean and variance for net-electric charge
and net-proton number fluctuations measured by the STAR and PHENIX Collaborations.
For the STAR electric charge data \cite{STARQ} we show results obtained by normalizing 
with net-proton results from the published STAR0.8 data set
\cite{STAR08} (triangles)
as well as the preliminary STAR2.0 data set \cite{STAR20} (circles). 
The experimental data is compared to QCD predictions
for two values of $\kappa_2^f$ (see text).
The electric charge results obtained by PHENIX \cite{PHENIXQ}
have been normalized by using the STAR2.0 data set on net-proton fluctuations (boxes).
For orientation the upper x-axis shows $\sqrt{s_{NN}}$ energies of the RHIC beam energy 
scan with labels put at the values for $(M_P/\sigma_P)^2$ corresponding to the
STAR2.0 data set. Errors on  $(M_P/\sigma_P)^2$ are not visible as they are smaller than
the size of the symbols.
\vspace{-0.5cm}
}
\label{fig:R12QBdata}
\end{center}
\end{figure}

A determination
of $\Sigma_r^{QB}$ and $R_{12}^B$ at large $\sqrt{s_{_{NN}}}$ 
suffices to fix the freeze-out temperature $T_{f,0}$ and the 
quadratic correction $c_{12}$.
Combining this with lattice QCD results on $c_{12}^0$ and $D_{12}$ 
allows to extract the curvature, $\kappa_2^f$, of the freeze-out 
line.  However,
in heavy ion collisions $\Sigma_r^{QB}$ and $R_{12}^B$ are not
directly accessible.  Net-proton rather than net
baryon numbers, i.e. 
ratios like $R_{12}^P$ and $\Sigma_r^{QP}$ rather than 
$R_{12}^B$ and $\Sigma_r^{QB}$, are measured. The STAR Collaboration  
obtained $R_{12}^P$ in the transverse momentum interval
$0.4~{\rm GeV}\le p_t\le p_t^{max}$ with $p_t^{max}=0.8~{\rm GeV}$ 
(STAR0.8) \cite{STAR08}.
The $p_t$-range has recently been extended
and preliminary results up to $p_t^{max}=2.0~{\rm GeV}$ (STAR2.0) 
\cite{STAR20} have been presented.
The ratio of net-electric charge fluctuations, $R_{12}^Q$,
has been measured by STAR \cite{STARQ} and PHENIX \cite{PHENIXQ}
in the interval $p_t^{min}\le p_t\le 2.0~{\rm GeV}$
with $p_t^{min}=0.2$~GeV and $p_t^{min}=0.3$~GeV\footnote{In addition to this different coverage of the $p_t$
range for charged particles the measurements differ in the 
coverage of azimuthal angle. Furthermore STAR has eliminated all
protons and anti-protons with $p_t\le 0.4$~GeV from their data sample.},
respectively.  
Fig.~\ref{fig:R12QBdata} 
shows  results for 
$\Sigma_r^{QP}$ versus $\left( R_{12}^{P}\right)^2$. In the case of the PHENIX
data for $R_{12}^Q$, we used the STAR2.0 data set to construct
$\Sigma_r^{QP}$.

Before entering into details of 
the analysis of these data using lattice QCD calculations
for $R_{12}^B$ and $R_{12}^Q$ we need to discuss systematic effects that
arise in equilibrium thermodynamics\footnote{We refrain here from a 
discussion of possible non-equilibrium effects \cite{Mukherjee:2015swa} 
or effects arising
from additional inelastic scatterings after freeze-out \cite{Kitazawa:2011wh}. 
This may play a role in the quantitative analysis of charge fluctuations,
but goes beyond the equilibrium framework we try to establish here.} 
because net-proton rather than net baryon numbers 
are measured in experiments
and because only data from a limited region in momentum space is available.  
The influence of low and high momentum cuts on charge fluctuations
has been analyzed in HRG models \cite{Nu,Morita}. The most important 
effects arise from a non-zero $p_t^{min}$ which most drastically influences
the pion contributions to net-electric charge fluctuations.
Implementing the $p_t$-cuts STAR has used for protons and other charged
particles \cite{STARQ} in a  HRG model calculation suggests that 
$R_{12}^Q$ is overestimated by about 5\% while
the larger $p_t^{min}=0.3$~GeV used by PHENIX \cite{PHENIXQ} amounts to
an increase of 20\% \cite{Morita}. This explains about 40\% of the difference 
in $R_{12}^Q$ seen by STAR and PHENIX. 
It is conceivable that the remainder arises from the
small azimuthal coverage in the PHENIX experiment which further reduces
the acceptance of charged particles \cite{Bzdak:2012ab}. In order to
get better controll over these effects a more detailed
experimental study of systematics of arising from non-zero $p_t$ cuts
and cuts in the rapidity coverage will be needed.

With proton number not being a conserved quantity \cite{Kitazawa:2011wh}, 
the ratio $R_{12}^{P}$
clearly has no meaning in the high temperature plasma phase of QCD.
At chemical freeze-out, however, when inelastic interactions are no
longer of relevance, the net proton number may be considered to be a
well defined concept. Still $R_{12}^{P}$ and $R_{12}^{B}$ will differ
in equilibrium thermodynamics.
The difference can be estimated in a HRG model calculation 
where $R_{12}^P=\tanh (\hmu_B+\hmu_Q)$, independent
of the value $\hmu_S$, while
$R_{12}^B$ explicitly depends on $\hmu_S$  (Eq.~\ref{chiX}).
For our NLO analysis of cumulant ratios it suffices to determine the LO
relation between $R_{12}^P$ and $R_{12}^B$. Using Taylor expansions for
both, this obviously yields $R_{12}^B/R_{12}^P= R_{12}^{B,1}$, which is shown
in the middle panel of Fig.~\ref{fig:chi2QB}.

The above discussion suggests that in the STAR  measurements
the systematic errors of
$R_{12}^Q$ and $R_{12}^B$ are of similar magnitude
and tend to cancel to a large extent in the ratio $\Sigma_r^{QB}$; i.e.
the ratio $\Sigma_r^{QP}$ indeed seems to be a good
proxy for $\Sigma_r^{QB}$  while for the PHENIX set-up 
it overestimates $\Sigma_r^{QB}$ by at least 10\%.

We have fitted the three data sets corresponding to the published
STAR data (STAR0.8), the preliminary STAR data (STAR2.0) and the
PHENIX data on net-charge fluctuations normalized to the 
STAR data on net-proton fluctuations (PHENIX/STAR2.0) 
using the ansatz given in Eq.~\ref{relation}. 
These data are shown in Fig.~\ref{fig:R12QBdata}.
For each of these three
data sets two different fit ranges have been chosen,
$R_{12}^P\le 0.6$ and 0.8, 
which correspond to data taken
at $\sqrt{s_{_{NN}}}\ge 39$~GeV and 27~GeV, respectively. This determines 
the intercept $a_{12}$ and the curvature parameter $c_{12}$ given in
Table~\ref{tab:fit}.  Differences arising from the two fit ranges 
have been added as systematic error in the error analysis of $a_{12}$
and $c_{12}$. From the intercept at $R_{12}^{P/B}=0$ the freeze-out 
temperature $T_{f,0}$ is obtained.
Once $T_{f,0}$ is fixed this way the NLO expansion coefficients $c_{12}^0$ 
and  $D_{12}$ are also fixed (see Table~\ref{tab:fit}) and we
obtain QCD predictions for  $\Sigma_r^{QB}$ with
$\kappa_2^f$ as the sole free parameter.
In Fig.~\ref{fig:R12QBdata} we show $\Sigma_r^{QB}$ as a 
function of $\left( R_{12}^P\right)^2$ for $\kappa_2^f=0$
(colored bands). Note that any value $\kappa_2^f >0$ will result in a 
weaker dependence of  $\Sigma_r^{QB}$ on $R_{12}^P$.
As an illustration we also show the result
for $\Sigma_r^{QB}$ at $T_{f,0}$ and $\kappa_2^f=0.02$.
as black lines.

\section{Curvature of the freeze-out line}
We now can discuss constraints for the curvature coefficient $\kappa_2^f$
resulting from the measured cumulant ratios $\Sigma_r^{QB}$. We consider 
the STAR data where cumulant ratios of net-electric charge as well as net-proton
number fluctuations have been measured. As discussed earlier we consider $\Sigma_r^{QP}$
to be a good approximation for the ratio $\Sigma_r^{QB}$. The  
difference between $R_{12}^P$ and $R_{12}^B$ can be corrected for by using the
HRG model motivated correction, $R_{12}^P= R_{12}^B/R_{12}^{B,1}$. This is
appropriate for our NLO approximation and simply amounts to a rescaling of 
the abscissa in Fig.~\ref{fig:R12QBdata}. While the intercept $a_{12}$
in quadratic fits is not influenced by such a rescaling, $c_{12}$ increases
by a factor $(R_{12}^{B,1})^{-2}$, i.e. by about 20\% in the relevant
temperature range.  From Eq.~\ref{c12a} it is obvious that
this will decrease the estimate for $\kappa_2^f$. Analyzing 
the uncorrected STAR data thus will put an upper bound on $\kappa_2^f$.

\begin{table}
\begin{center}
\begin{tabular}{|c|c|c|c|}
\hline
~&STAR0.8&STAR2.0&PHENIX/STAR2.0 \\
\hline
$a_{12}$ & 0.079(3)&0.087(2)& 0.110(9) \\
$c_{12}$ & 0.858(101)&0.329(74) & 0.559(352) \\
\hline
$T_{f,0}$~[MeV] & 145(2)& 147(2) & 155(4)   \\
$c_{12}^{0}(T_{f,0})$ & 0.343(31)&0.326(32)  & 0.265(52) \\
$D_{12}(T_{f,0})$     & 7.04(44) & 6.62(36)  & 5.27(78)  \\
\hline
$\kappa_2^f$   & -0.073(16) &  -0.001(12) & -0.056(67)  \\
\hline
\end{tabular}
\end{center}
\caption{Parameters of a quadratic fit to the STAR data
on the ratio $\Sigma_r^{QP}$ and the combination of PHENIX
data on $R_{12}^Q$ and STAR data on $R_{12}^P$.
The third to fifth row give the freeze-out temperature 
$T_{f,0}$, $c_{12}^0$ and $D_{12}$ at fixed $T_{f,0}$. 
The last row gives the curvature coefficients $\kappa_2^f$
obtained from Eq.~\ref{c12a}. In the fits none of the corrections
discussed in the text have been taken into account.
}
\label{tab:fit}
\end{table}
 
Results for $c_{12}$ from quadratic fits to the two STAR data sets 
are given in Table~\ref{tab:fit}. 
We notice first that  $c_{12}$  extracted from
the published STAR data \cite{STARQ,STAR08} is about a factor 2 larger
than the lattice QCD result for $c_{12}^0(T_{f,0})$. 
Since $D_{12}$ is positive, this corresponds 
to negative values for $\kappa_2^f$ 
as observed also in a HRG model analysis \cite{Ratti}, i.e.
the STAR data on proton fluctuations taken in the range 
$0.4 {\rm GeV} \le p_t\le 0.8 {\rm GeV}$ \cite{STAR08} are only
compatible with a negative curvature coefficient $\kappa_2^f$.
However, the still
preliminary STAR data taken in the larger $p_t$ interval \cite{STAR20}, 
have a much smaller slope which is consistent with $c_{12}^0(T_{f,0})$
within statistical errors. 
This gives an upper bound on the curvature of the freeze-out line,
\begin{equation} 
\kappa_2^f < 0.011 \; . 
\end{equation}
Taking into account the HRG motivated correction for
replacing $R_{12}^P$ by $R_{12}^B$ reduces  $\kappa_2^f$ and
makes the estimate for the upper bound compatible with zero,
$\kappa_2^f=-0.012(15)$ for the STAR2.0 data set, i.e.
the existing data for $\Sigma_r^{QP}$ favor a small or even vanishing
curvature of the freeze-out line at large $\sqrt{s_{_{NN}}}$ as it is the 
case for the phenomenological parametrization given in Ref.~\cite{Andronic:2005yp}.

Let us finally compare the result obtained for $\kappa_2^f$  with 
the curvature coefficient $\kappa_2^B>0$ of the QCD transition line, 
\begin{equation}
T_c(\mu_B) = T_{c,0} \left( 1 - \kappa_2^B \hmu_B^2  + {\cal O} (\hmu_B^4)\right)\; ,
\label{Tc} 
\end{equation}
where $T_{c,0}$ denotes the transition temperature
at $\mu_B=0$ \cite{Aoki:2009sc,Tc}.
The above bound on $\kappa_2^f$ is consistent with determinations
of $\kappa_2^B$ based on expansions of $T_c(\mu_B)$ around $\mu_B=0$
where the curvature coefficient is determined as a Taylor expansion coefficient
evaluated at $\mu_B=0$. This gave  $\kappa_2^B\simeq 0.007$ 
\cite{Kaczmarek:2011zz,Endrodi:2011gv}, which is 
about a factor two smaller than recent results for $\kappa_2^B$
\cite{Bonati:2015bha,Bellwied:2015rza,Cea:2015cya} that
are based on lattice QCD calculations performed with large non-zero imaginary 
chemical potentials, corresponding to $\mu_B/T_{c,0} \simeq (1-3)$. 
These calculations yield values $\kappa_2^B\simeq (0.015-0.02)$. 

\section{Conclusions}

We provided a framework that allows to 
determine the curvature of the freeze-out line through a direct comparison
between experimental data for mean and variance of net-electric charge and 
net-proton number fluctuations with lattice QCD calculations of cumulant
ratios. We found the curvature of the freeze-out line to be small. It is 
consistent with the curvature of the QCD crossover line. At least for beam 
energies $\sqrt{s_{_{NN}}}\ge 27$~GeV our study suggests that freeze-out 
happens close to the crossover transition line.  

We have addressed some difficulties that arise when comparing 
lattice QCD calculations of conserved charge fluctuations with
experimental data on cumulants of net-proton and net-charge fluctuations,
although at present it is difficulty to provide a complete quantitative
approach for this.
We have pointed out some obvious differences between net-proton and 
net-baryon number fluctuations that are present already in equilibrium 
thermodynamics. We also addressed the question on the influence of 
transverse momentum cuts on the experimental data. Clearly control
over these effects needs to be improved in future work. 
It also is known that cumulants of conserved charge fluctuations
are sensitive to the width of the rapidity window covered by the
experiments \cite{ALICE,Sakaida}. This dependence is more significant
for net-electric
charge fluctuations than for net-proton number fluctuations.
Increasing the rapidity window will decrease $\sigma_Q$ and thus
will lead to an increase of $\Sigma_r$. This will lead to larger
values for $T_{f,0}$. However, at present these effects
are difficult to quantify. 

Finally we note that throughout our analysis we assumed that
the net-charge to net-baryon number ratio, $r$, is determined by
the corresponding ratio present in the incident beams, $r\simeq 0.4$.
This ratio clearly will fluctuate in a fixed acceptance range covered
in an experiment and one may argue that
this ratio also shifts towards the isospin symmetric limit $r=0.5$
for high beam energies. Clearly, at present beam energies $r<0.5$,
as the net-charge expectation value would vanish in the isospin
symmetric limit. In a more refined analysis one may, however,
take into account effects arising from changes in the value of
$r$ which may also be beam energy dependent. In general, 
increasing $r$ results in a decrease of $T_{f,0}$ and an
increase of $\kappa_2^f$. Compared to other
uncertainties, however, this effect is small. 
For the preliminary STAR data set
(STAR2.0) we find, for instance, that using $r=0.45$ decreases
the $T_{f,0}$ by 3~MeV and shifts the curvature coefficient to
a slightly positive value, $\kappa_2^f=0.004$.

{\it Acknowledgements:} 
This research used resources of the John von Neuman Center in J\"ulich, 
Germany, made available through a PRACE grant. Computations at
the Oak Ridge Leadership Computing Facility, 
which is a DOE Office of Science User Facility supported under Contract 
DE-AC05-00OR22725, were made possible through an ALCC grant.
This work has been partially supported through
the U.S. Department of Energy under Contract No. DE-SC0012704,
and the Bundes\-ministerium f\"ur Bildung und Forschung (BMBF)
under grant no. 05P15PBCAA.

\end{document}